\documentclass[preprint]{ptephy}
\preprintnumber{}
\begin{document}

% %%%%%ptep

\title{The phase synchronization of an axion and a superconductor}

\author{\name{\fname{Hideto} \surname{Manjo}}{1}, \name{\fname{Koichiro} \surname{Kobayashi}}{2}, \name{\fname{Kiyoshi} \surname{Shiraishi}}{1}}

\address{
\affil{1}{Yamaguchi University, Yamaguchi-shi, Yamaguchi 753-8512, Japan} 
\affil{2}{Kyushu Institute of Technology, 680-4 Kawazu, Iizuka-shi, Fukuoka 820-8502, Japan} 
\email{s004wa@yamaguchi-u.ac.jp}}

\begin{abstract}
The effects of the axion field have been widely studied in theoretical physics, particularly in particle physics. 
Considering the phase synchronization and the mean free path of the axion, the bulk of the phase coherent superconductor is regarded as the weak link region of the Josephson junction.
It is expected that the axion mass influences the London penetration depth.
There is a slight possibility of detecting this effect because the effect becomes more significant in superconductor with a low carrier density $n_s$.
%Limits on our results imply $ m_a c^2 \lesssim 1 \ \mathrm{MeV}$ at least.
The differences due to the choice of axion model and the axion mass are discussed.
\end{abstract}

\subjectindex{B04,B50,C15,I61}

% %%%%%%%%%%%%%%%%%%%%%%%%%%%%%%%%%%%%%%%%%

\maketitle
%%%%%%%%%%%%%%%%%%%%%%%%%%%%%%%%%%%%%%%%%%%%%%%%%%%

\section{Introduction}
	Many authors have proposed various theories of massive photons.
In this paper, the topologically massive model associated with the axion field is studied.
The axion electrodynamics is an extension of Maxwell's electromagnetic theory that includes the dynamical axion term.
The presence of the dynamical axion term means that the photon becomes topologically massive in the axion electrodynamics. 

	The axion term is also called the Chern-Simon (CS) term because of its origin. 
The chiral magnetic effect (CME) \cite{kharzeev2010topologically,PhysRevD.78.074033,PTPS.193.15} is a well-known topologically induced electromagnetic effect in the presence of the time-dependent CS term.
The effect of the CS term has been reported in the literature \cite{wilczek1987two,burgess1993classical,lehnert2004vcerenkov,alves2010repulsive}.
In this connection, it is probable that the CS term affects the properties of matter \cite{ooguri2012instability}.
More detailed studies \cite{randjbar12304844,grigorio2013chern,shiozaki2014} have addressed the role of the CS term or axion term in superconductors.

	It was reported recently that the axion mass can be estimated using resonant Josephson junctions, assuming a time-dependent axion field \cite{beck2014axion,beck2013possible}.
These studies reported that the observed Shapiro step anomalies in all four experiments consistently point toward an axion mass of $ \left( 110 \pm 2 \right) \ \mu  \mathrm{eV}$.
As the author of \cite{beck2013possible,beck2014axion} pointed out, this result for the axion mass also needs to be examined from other viewpoints or experimentally.

	In present paper, it has been shown that the relation of $\dot{\theta}$ and axion mass $m_a$ in the bulk of superconductor from the beginning of the phase synchronizing condition and the London equations.
These findings suggest the axions penetrate deep inside of the superconductor and $\dot{\theta}$ enhances in the superconductor, and besides, it is probable that the measurement of the following modified London penetration depth allows the checking of presence of this phenomenon.  

	Among the various possible effects of the axion field, we focus on the London penetration depth.
Superconductors have perfect diamagnetism, which is called the Meissner effect.
Because of the Meissner effect, the magnetic field does not penetrate deep inside of the superconductor, and the depth is called the London penetration depth.
It is inferred from the presence of enhanced $\dot{\theta}$ in the superconductor that the London penetration depth is due to the axion mass $m_a$. 
This paper presents the simple classical results for the phase synchronization of bulk of superconductor and the London penetration depth of a Type-I superconductor using the electromagnetic field theory, including the time-dependent axion field.

\section{Dynamics of photons and axions}
	The Lagrangian density for the axion electrodynamics (Maxwell--Chern--Simons equations \cite{wilczek1987two,kharzeev2010topologically,PTPS.193.15}) and the axions dynamics is written as the sum of the Lagrangian densities for the classical electromagnetism, the axion's two-photon interaction and the axions dynamical term \cite{sikivie2007,long1995}:
\begin{equation} 
\mathcal{L}_{a} = -\frac{1}{4} F_{\mu \nu} F^{\mu \nu } + \theta \frac{g_\gamma e^2 }{16 \pi^2} F_{\mu \nu} \tilde{F}^{\mu \nu} + \frac{1}{2} f^2_a \partial_\mu \theta \partial^\mu \theta - \frac{1}{2} f^2_a m^2_a \theta^2 - j^\mu A_\mu, \label{eq1}
\end{equation}
where $g_\gamma$ is a model-dependent coupling constant having a value of $g_\gamma =-0.97$ for KSVZ axions \cite{kim1979weak,shifman1980can} or $g_\gamma =0.36$ for DFSZ axions \cite{zhitnitskii1980possible,dine1981simple}.
$\theta(\mathbf{x})=\phi_a (\mathbf{x})/f_a$ is the misalignment angle of the axion field $\phi_a(\mathbf{x})$, and $f_a$ is the axion decay constant;
$-e$ is the charge of an electron.
The speed of light, the vacuum permittivity constant, the vacuum permeability constant, the reduced Planck constant are defined as $c=1$, $\varepsilon_0=1$, $\mu_0=1$ and $\hbar=1$, respectively. 
The gauge and Lorentz invariance cannot rule out a second term including $\theta(\mathbf{x})$. 
In other words, it is possible to allow a slight $\theta(\mathbf{x})$ dependence.
Moreover, the behavior of this dynamical $\theta(\mathbf{x})$ is worth considering.
%%%%%%%%%%%%%%%%%%%%%%%%%%%%%%%%%%%%%%%%%%%%
% It is a straightforward calculation to deduce the equations of motion from the Lagrangian \eqref{eq1}:
% \begin{eqnarray}
% \nabla \cdot \mathbf{E} &=& \rho - \frac{g_\gamma e^2 }{4 \pi^2} (\nabla \theta) \cdot \mathbf{B}, \label{eq2}\\
% \nabla \times \mathbf{B} - \partial_t \mathbf{E} &=& \mathbf{j} - \frac{g_\gamma e^2}{4 \pi^2} \left[ (\partial_t \theta) \mathbf{B} - (\nabla \theta) \times \mathbf{E} \right], \label{eq3}\\
% \partial^2_t{\theta} - \nabla^2 \theta + m^2_a \theta &=& -\frac{g_\gamma e^2}{4 \pi^2 f^2_a} \mathbf{E} \cdot \mathbf{B} \label{axion_grand_eq},
% \end{eqnarray}
% The other two expressions in Maxwell's equation do not change ($\nabla \cdot \mathbf{B}=0,\nabla \times \mathbf{E} = - \frac{\partial \mathbf{B}}{\partial t}$).
% In Eqs. \eqref{eq2}, \eqref{eq3} and \eqref{axion_grand_eq}, assuming that $\theta$ depends only on the $t$ coordinate, the differential term with respect to the space coordinate, $\nabla \theta$, can be eliminated.
% Thus, expressions \eqref{eq2} to \eqref{axion_grand_eq} can be written as
% \begin{eqnarray}
% \nabla \cdot \mathbf{E} &=& \rho,  \label{eq4}\\
% \nabla \times \mathbf{B} - \partial_t \mathbf{E} &=& \mathbf{j} - \alpha \dot{\theta} \mathbf{B},  \label{eq5}\\
% \ddot{\theta} + m^2_a \theta &=& -\frac{\alpha}{f^2_a} \mathbf{E} \cdot \mathbf{B} \label{axion_time_eom}
% \end{eqnarray}
% where $\dot{\theta} \equiv \partial_t \theta$ and $\alpha = (g_\gamma e^2 )/(4 \pi^2)$.
%%%%%%%%%%%%%%%%%%%%%%%%%%%%%%%%%%%%%%%%%%%%%
It is a straightforward calculation to deduce the equations of motion from the Lagrangian \eqref{eq1}:
\begin{eqnarray}
\nabla \cdot \mathbf{E} &=& \rho,  \label{eq4}\\
\nabla \times \mathbf{B} - \partial_t \mathbf{E} &=& \mathbf{j} - \alpha \dot{\theta} \mathbf{B},  \label{eq5}\\
\ddot{\theta} + m^2_a \theta &=& -\frac{\alpha}{f^2_a} \mathbf{E} \cdot \mathbf{B}. \label{axion_time_eom}
\end{eqnarray}
where assuming that $\theta$ depends only on the $t$ coordinate, the differential term with respect to the space coordinate, $\nabla \theta$, can be eliminated, and $\dot{\theta} \equiv \partial_t \theta$ and $\alpha = (g_\gamma e^2 )/(4 \pi^2)$ are defined.
The other two expressions in Maxwell's equation do not change ($\nabla \cdot \mathbf{B}=0,\nabla \times \mathbf{E} = - \frac{\partial \mathbf{B}}{\partial t}$).
The second term on the right-hand side in Eq. \eqref{eq5} represents the current, and this current is called $\mathbf{j}_{\mathrm{CME}}$ (the chiral magnetic current) \cite{PTPS.193.15}.
The purpose of this paper is to estimate the effect of this term.

\section{London penetration depth}
	The London equations describe the Meissner effect phenomenologically \cite{london1935}.
From the London equations, the magnetic field is written as a rotation of the current,
\begin{equation}
\mathbf{B} = -\frac{m_e}{n_s e^2 } \left( \nabla \times \mathbf{j} \right), \label{Meissner_cond}
\end{equation}
and Eq. \eqref{Meissner_cond} represents perfect diamagnetism.
% \footnote{This equation is also derived from the London moment $\mathbf{B}=(2m \boldsymbol{\omega} )/(e)$, where $\boldsymbol{\omega}$ is the angular velocity of superconducting electrons inside the superconductor. The London moment is the magnetic moment acquired by a rotating superconductor and was predicted by Becker, Heller, and Sauter \cite{becker1933stromverteilung} in 1933, just before the Meissner effect was discovered. Reference \cite{hirsch2014london} gives more details on these matters. The London moment was observed by the GP-B experiment \cite{hipkins1996measurement}}
In the axion electrodynamics, the additional magnetic field on the right-hand side of Eq. \eqref{eq5} is in the same direction as the current.
Namely, it is necessary to calculate the London penetration depth under slightly unusual conditions.
This paper shows a method of deriving the modified London penetration depth under simple assumptions.

	Let us now apply a rotation to both sides of Eq. \eqref{eq5} and substitute the perfect magnetism \eqref{Meissner_cond} in the equation. The magnetic field equation is written as
\begin{equation}
 \nabla^2 \mathbf{B} = \beta \mathbf{B} + \alpha \dot{\theta} \nabla \times \mathbf{B}, \label{magnetic_eq}
\end{equation}
where  $\beta=(n_s e^2)/(m_e)$, and $\partial_t \mathbf{E}=0$ are defined.
Note that the second term of Eq. \eqref{magnetic_eq} depends on $\dot{\theta}$. 

	Next, the following two conditions apply.
First, the superconductor is placed in the region $x>0$.
Second, the magnetic field and current depend only on the $x$ direction.
Then, the magnetic field $\mathbf{B}$ and current $\mathbf{j}$ are expressed as $\mathbf{B}=B_y (x) \mathbf{e}_y +B_z (x) \mathbf{e}_z$ and $\mathbf{j} = j_y (x) \mathbf{e}_y +j_z (x) \mathbf{e}_z$, respectively, in Cartesian coordinates.
From Eq. \eqref{Meissner_cond}, the $y$ and  $z$ components of Eq. \eqref{magnetic_eq} are rewritten as
\begin{eqnarray}
\frac{\partial^2 B_y(x) }{\partial x^2} =  \beta B_y (x) - \alpha \dot{\theta} \frac{\partial}{\partial x} B_z (x), \label{B_eq_y}\\
\frac{\partial^2 B_z(x) }{\partial x^2} =  \beta B_z (x) + \alpha \dot{\theta} \frac{\partial}{\partial x} B_y (x). \label{B_eq_z}   
\end{eqnarray}

	Here, we introduce the ratio of $B_y$ and $B_z$:
\begin{equation}
\frac{B_z (x)}{B_y (x)} = \tan \eta \label{B_ratio}.
\end{equation} 
This equation states that the magnetic field decreases, whereas the magnetic field ratio \eqref{B_ratio} inside the superconductor is maintained.
Substituting Eq. \eqref{B_ratio} into Eqs. \eqref{B_eq_y} and \eqref{B_eq_z}, we obtain
\begin{equation}
\beta  B_y (x)=\frac{\partial^2}{\partial x^2} B_y (x)- 2 \Gamma(\eta) \alpha \dot{\theta}   \frac{\partial}{\partial x} B_y (x),\label{B_modified_eq}
\end{equation}
where the function $2 \Gamma(\eta) = (1- \tan \eta)/(1+ \tan \eta)$ is defined as that composed of the ratio variable $\eta$.
It is important to note that the magnetic field becomes zero deep inside of the superconductor to estimate the London penetration depth. 
This boundary condition is $\lim_{\mathbf{x} \to \infty} \mathbf{B} = 0$. 
Assuming $\lim_{x \to 0}B_y(x) = B_{y0}$, Eq. \eqref{B_modified_eq} yields a simple magnetic solution:
\begin{eqnarray}
B_y (x) &=&  B_{y0} e^{-\frac{x}{\lambda_a}},\\
\frac{1}{\lambda_a}&=&  \sqrt{\beta + \Gamma(\eta)^2 \alpha ^2 \dot{\theta}^2 }-  \Gamma(\eta) \alpha \dot{\theta}.  \label{Bsol}
\end{eqnarray}
When $\dot{\theta} = 0$, this equation corresponds to the original London penetration depth $\lambda_L$.

\section{Phase synchronization of the superconductor}
	In the literature \cite{beck2013possible}, the author states that the axions into weak link region of the Josephson junction immediately decays.
In present paper, we consider a bulk of phase coherent superconductor that lies in $\theta$ space, instead of considering the Josephson junction.
Assuming the superconductor synchronize with the axion field, it is found that the axion phase $\dot{\theta}$ enhances in the superconductor.
Our result is base on a new point of view, i.e., the London equations.
Let us now consider the phase synchronization of the superconductor and axion decay.

	The superconductor is under the status of Bose-Einstein condensates (BEC), that implies the the phase $\varphi$ of the wave function $\Psi_\mathrm{sc}= |\Psi_\mathrm{sc}| e^{i \varphi}$ is synchronized on the bulk of the superconductor.
Also it means that the superconductor is the microscopic quantum object.
Now, we define that the wave function of the exterior of the superconductor $\Psi_\mathrm{ext}= |\Psi_\mathrm{ext}| e^{i \theta}$ and the interior of the superconductor $\Psi_\mathrm{int}= |\Psi_\mathrm{int}| e^{i \varphi}$
where the phase difference of both the wave functions is $\delta = \theta -\varphi$.
If the incoming axions enter the superconductor, then the region around axions is the different phase to the external $\theta$ vacuum space.
Therefore, the phase difference $\delta$ emerges, and produces the weak link like region of the Josephson junction in the superconductor.

	Grant that this axion generated region is considered as weak link region of the Josephson junction, the following equations holds,
\begin{equation}
\frac{\mathrm{d} \delta} {\mathrm{d} t} = 2e V. \label{j2}
\end{equation}
where $V$ is the difference of a voltage of the inside of superconductor and the outside of the superconductor. 
From this relation, it is found that the time derivatives of the phase satisfy the relation $\dot{\delta} = \dot{\theta}-\dot{\varphi}=2eV$.
Assume that the initial condition of the phase of the superconductor $\varphi=\varphi_0$,
we get the phase synchronization condition $\dot{\delta} = \dot{\theta} = 2eV$, that equivalents to the Beck's phase synchronization condition \cite{beck2013possible} about the weak link region.

	The main key for describing the enhancement of the phase $\dot{\theta}$ is the relation of the electromagnetic field and the current, i.e., the London equations.
The electric field in the superconductor obeys the London equation for the electric field:
\begin{equation}
\mathbf{E} = \frac{1}{\beta} \frac{ \mathrm{d} \mathbf{J}}{\mathrm{d} t} + \nabla \rho. \label{london_eq2}
\end{equation}
This equation implies that the time derivative of the density of the current produces the electric field $\mathbf{E}$, and the static current does not produce the electric field in the superconductor.
From the phase synchronization condition $\dot{\delta} = \dot{\theta}$, we get the relation of $\theta$ and $J$, that are $\dot{\theta} = (2e d \dot{J})/( \beta)$, $\ddot{\theta} = (2e d \ddot{J})/( \beta)$, and $\theta = (2ed J)/( \beta) + c_1$ where $\nabla \rho = 0$, $V=Ed$.
If $J = 0$ and $\delta=0$ satisfies, then $\delta = c_1 - \varphi_0 = 0$, this means $c_1 = \varphi_0$.
Moreover, the magnetic field is written as
\begin{equation}
B =  \frac{1}{2ed \dot{J}} \frac{ \beta}{\alpha} \left( \frac{1}{\beta} \ddot{J} + J \right), \label{axion_b_eq}
\end{equation}
from Eq.\eqref{eq5} and Eq.\eqref{london_eq2} where the electric field $E$ is in the same direction as the the magnetic field $B$.
Substitute these $\dot{\theta}$, $\ddot{\theta}$,  Eq.\eqref{london_eq2} and Eq.\eqref{axion_b_eq} to Eq.\eqref{axion_time_eom}, the equation of motion of axion is rewritten as the current equation:
\begin{equation}
 \left[ 1 + \frac{1}{f^2_a}\left(\frac{1}{2ed} \right)^2  \right] \ddot{J} + \left[ m^2_a +  \frac{\beta}{f^2_a} \left(\frac{1}{2ed} \right)^2   \right] J = 0, \label{current_eq}
\end{equation}
where $\varphi_0 \ll 1$ is used.
In the later, it is found that the second term and the forth term in Eq.\eqref{current_eq} are negligible.
However, we now examine this equation, because of the order of $d$ is unknown.
Solving Eq.\eqref{current_eq}, we now get the oscillation solution
$J = J_0 \sin \left( \omega t + \delta_0 \right)$
, and $\omega$ is written as 
\begin{equation}
\omega^2 = \frac{ \left[ 
m^2_a + \frac{\beta}{4 e^2 d^2 f^2_a} \right]
}{\left[  1  + \frac{1}{4 e^2 d^2 f^2_a} \right]} \label{omega_velocity},
\end{equation}
about the current $J$ where $J_0$ and $\delta_0$ are constant.
This equation means that the current is oscillation in the superconductor.
Here, if this current $J$ is regard as the Josephson junction current, we obtain $\delta = \omega t + \delta_0$ and  $\dot{\delta} = \partial_t (\omega t + \delta_0)= \omega = 2eV$.
This result implies that, in the region of surface to length $d$, the current obey 
\begin{eqnarray}
J = \frac{\beta}{2ed} \left(  \omega t + \delta_0 \right) \label{liner_current},
\end{eqnarray}
since the relation of $\theta$ and $J$, and $\theta= \delta$.
	Using this relation, the entering axion generated magnetic field are written as
\begin{equation}
B=\frac{1}{2 e d} \frac{\beta}{\alpha} t = \frac{2\pi^2}{g_\gamma} \frac{n_s}{e m_e d} t, \label{StrongB}
\end{equation}
in the weak link like region from Eq.\eqref{london_eq2} and Eq.\eqref{axion_b_eq} for $J|_{t=0} = \delta_0 = 0$.
%Note that Eq.\eqref{omega_voltage_relation} shows $2eV=\omega$, which implies the energy of the angular velocity of the junction like region $\hbar \omega$ is equals to the electric energy of an electron pair $2eV$.

	Next, the length of the axions decay is estimated.
We consider the situation that an axion is placed the surface of the superconductor at $t=0$ and has the velocity $v_a$.
From Eq.\eqref{StrongB}, the time average of magnetic field in the weak link like region becomes
\begin{equation}
\bar{B} = \frac{ \pi^2 \hbar^2}{g_\gamma} \frac{n_s}{e m_e d} T =  \frac{ \pi^2 \hbar^2}{g_\gamma} \frac{n_s}{e m_e v_a},
\end{equation}
in SI units where $d=v_a T$ is used.
Putting typical values for the superconducting electron density $n_s=10^{28} \ \mathrm{m}^{-3}$, the axion velocity $v_a=2.3 \times 10^5 \ \mathrm{m/s}$ and the axion mass $ m_a c^2 = \left( 110 \pm 2  \right) \ \mu \mathrm{eV}$.
As the numerical example, the magnetic field $\bar{B} = -3.4 \times 10^4 \ \mathrm{T}$ for the KSVZ axion is found.
The Primakoff effect is estimated by using this result.
The probability of axion decay is given by following equation \cite{sikivie2007}:
\begin{equation}
P_{a \to \gamma}  = \frac{1}{4 v_a} \left(g \bar{B} L \right)^2  \left( \frac{\sin \frac{qL}{2}}{\frac{qL}{2}} \right)^2,
\end{equation}
where $q$ is axion-photon momentum transfer, $L$ is an axion flying distance, and $g= (g_\gamma e^2)/(4 \pi^2 f_a)= \alpha/f_a$.
In $P_{a \to \gamma} =1$, an expression for the mean free path in the weak link like region is
\begin{equation}
L^2 =  \frac{64}{\mu_0 c^4 \hbar^3}  \frac{m^2_e f^2_a}{e^2 n^2_s} v^3_a,
\end{equation}
for $qL \ll 2 \hbar$ in SI units.
As the numerical example, we obtain the long distance $L= 10^{7} \ \mathrm{m}$.
Note that the distance is independent on the axion model since the mean free path $L$ does not include $g_\gamma$.

\begin{figure}[h]
\begin{center}
\includegraphics[width=70mm]{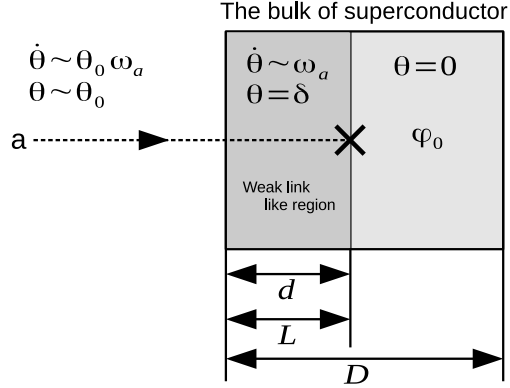}
\end{center}
\caption{The phase synchronization of the axion and the superconductor. The axion decays at $L$. \label{fig:one}}
\end{figure}

	Here, we consider that an entering axion decay at $L$ (See Fig.\ref{fig:one}).
In the region more deeper than distance $L$, it can be assumed that the phase $\theta$ that is propagated an axion does not exist, which implies that the weak link like region depth $d$ is equals to the mean free path $L$ for the superconductor that has a thickness $D>L$, namely $d=L$.
In that case, the angular velocity is rewritten as
\begin{equation}
\omega^2 = 
\frac{
\left[ 
\frac{m^2_a c^4}{\hbar^2} + c^3 \hbar^3  \frac{1}{\left(2 L f_a \right)^2} \frac{n_s}{m_e} 
\right]
}{
\left[
1  +  \frac{ c \hbar^3}{\mu_0} \frac{1}{\left(2 L f_a e \right)^2}
\right]},
\end{equation}
in SI units.
As the order estimation, the first term of above is $10^{22} \ \mathrm{s^{-2}}$, the second term of above is $10^{-40} \ \mathrm{s^{-2}}$, and the second term of below is $10^{-67}$.
Therefore, the second term of above and the second term of below are clearly negligible.
Note that if the superconductor has a thickness $D < L$, then the edges effect act on the angular velocity $\omega$ from \eqref{omega_velocity}.
However in this case, these edge effects is very tiny, still the second term of above and the second term of below have no more than $10^{-26} \ \mathrm{s^{-2}}$ and $10^{-53}$ for $D = 1 \ \mathrm{m}$.
Therefore, it is reasonable to suppose that $\omega \sim (m_a c^2)/(\hbar)$ in SI units is consistent.
% More basic detailed study of the superconductors also suggests similar phenomenon in this \cite{shiozaki2014}.  

	As the summary in this section, it is found that the bulk of superconductor for $D < L$ is regarded as the bulk of the weak link like region, and the axion mass $m_a$ is related to the angular velocity of the Josephson junction like current, that is to say
\begin{equation}
\omega = \frac{m_a c^2}{\hbar} = \dot{\delta} = \dot{\theta} = \frac{2eV}{\hbar},
\end{equation}
in SI units.

\section{Shapiro step of the axion}
\begin{figure}[h]
\begin{center}
\includegraphics[width=50mm]{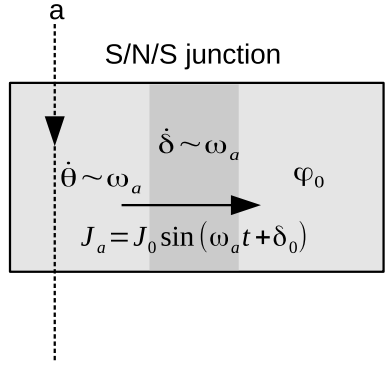}
\end{center}
\caption{S/N/S junction with an incoming axion. \label{fig:two}}
\end{figure}

	We consider the junction that is made of this synchronized superconductors and normal metal.
In this section, we use SI units.
If an axion enter the superconductor on the other hand either, and the phase of axion $\theta$ synchronize with the superconductor (See Fig.\ref{fig:two}). 
Then the phase difference $\dot{\delta} = \omega = (m_a c^2)/ \hbar$ as many as an axion emerges on the weak link region of normal metal, which make the super current of Josephson junction and the axion generated voltage $V_a=(m_a c^2)/(2e) = (\omega_a \hbar)/(2e)$.
So, the junction voltage bias is written as $V =  V_0 \pm (\omega_a \hbar)/(2e) $, which implies $\delta = \delta_0 + ( 2eV_0/\hbar \pm \omega_a )t $.
Inserting this $\delta$ into $J = J_0 \sin \delta$, this means a dc component only when $ 2eV_0 \pm \omega_a \hbar  = 0 $, i.e., when the dc voltage has the Shapiro step values $V_0 = (\mp \hbar \omega_a)/(2e)$.
Therefore, it is a possible that our scenario also describe the observed Shapiro step \cite{bae2008} and the unknown differential conductance peek \cite{hoffmann2004} in the S/N/S junction experiment.

\section{Discussion}
	The effect of the axion field about the London penetration depth is estimated from the results. 
Assume that the magnetic field ratio variable $\eta \to 0$, namely, $2 \Gamma \to 1$, the external magnetic field has only a $y$ component.
In SI units, the modified London penetration depth are rewritten as
\begin{equation}
 \frac{1}{ \lambda_a}=\sqrt{ \mu_0 \frac{n_s e^2}{m} + \frac{1}{4}
 \left( \frac{\mu_0}{\hbar} \frac{ {g_\gamma} e^2 \dot{\theta} }{4 \pi^2}\right)^2}  -\frac{\mu_0}{2 \hbar} \frac{g_\gamma e^2 \dot{\theta}  }{4 \pi^2}. \label{eq12}
\end{equation}

\begin{figure}[h]
\begin{center}
\includegraphics[width=137mm]{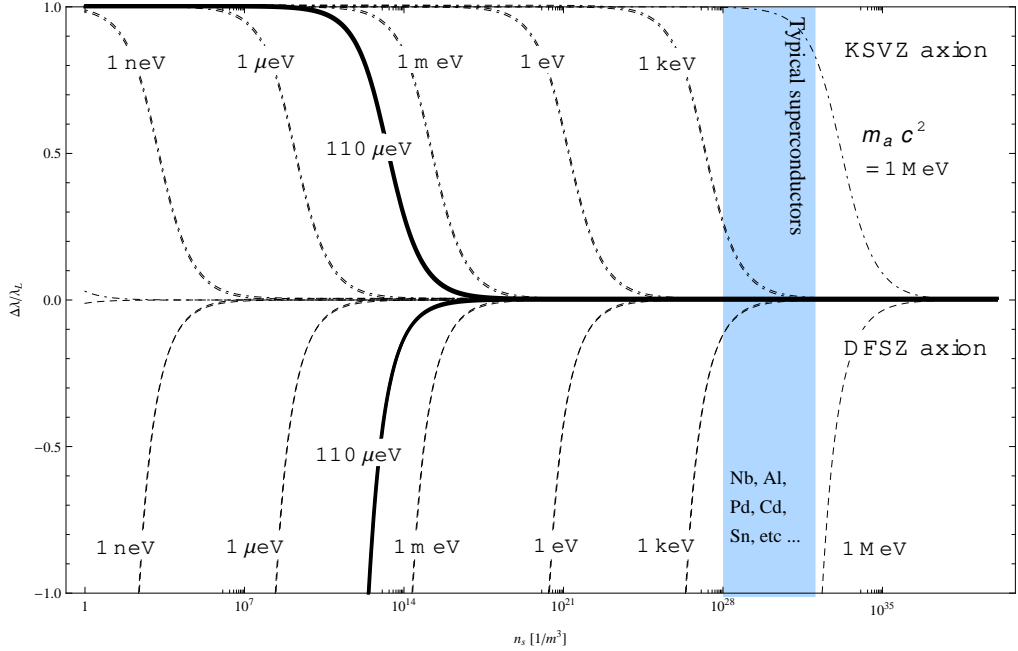}
\end{center}
\caption{$\Delta \lambda / \lambda_L$ for KSVZ axions and DFSZ axions, and the relationship between $n_s$ and $\Delta \lambda / \lambda_L$. The dash-dotted lines show various axion masses $m_a c^2 $; the thick line shows $ m_a c^2 = 110  \ \mu\mathrm{eV}$.
\label{fig:three}}
\end{figure}

	Next, the relationship between the axion mass $m_a$ and $\dot{\theta}$ is considered.
% If the incoming axions decay into microwave photons by above mechanism, the axion field behaves as in the weak link region of Josephson junctions at the surface of the superconductor.
% Then, from the literature \cite{beck2013possible,beck2014axion}, the linear solution $\theta(t) = \omega_a t +\mathrm{const}$ is possible.
The frequency, $\omega  =  m_a c^2 / \hbar =\dot{\delta} = \dot{\theta} $, is given by the axion mass.	
From the literature \cite{beck2014axion}, the axion mass is $ m_a c^2 = \left( 110 \pm 2 \right)  \ \mu\mathrm{eV}$, which implies $\dot{\theta} \sim 1.7 \times 10^{11} \ \mathrm{s^{-1}}$. Assume this value and the superconducting electron density of niobium $n_s\mathrm{(Nb)} \sim 1.5 \times 10^{28} \ \mathrm{m^{-3}}$, from the value of $\lambda_L$ in the literature \cite{gubin2005dependence}.
Then, the difference between the modified London penetration depth $\lambda_a$ and the original London penetration depth $\lambda_L$ is $\Delta \lambda = \lambda_L-\lambda_a$, where ${\Delta \lambda}_{\mathrm{KSVZ}} \sim 1.1 \times 10^{-15}\ \mathrm{m}$, and ${\Delta \lambda}_{\mathrm{DFSZ}} \sim -4.3 \times 10^{-16}  \ \mathrm{m}$.
It seems that the effect of the light axion on typical superconductors is insignificant. 
However, the effect of the axion for $ m_a c^2  > 1 \ \mathrm{MeV}$ is already visible (see Fig.\ref{fig:three}).

	As another possibility, the modified London penetration depth \eqref{eq12} is found to depend only on the density of the superconducting electrons $n_s$ except for $\dot{\theta}$ as a physical variable.
The axion field becomes significant if $n_s$ has a very low value.
Therefore, there is some possibility of detecting the effect of the light axion on superconductors having a low carrier density $n_s$.
If the axion mass is $ m_a c^2 = \left( 110 \pm 2  \right) \ \mu \mathrm{eV}$, it is expected that the effect of the axion mass in both the KSVZ and DFSZ models becomes prominent in the region $n_s< 10^{14} \ \mathrm{m^{-3}}$.
The sign of $\Delta \lambda$ is positive for KSVZ axions and negative for DFSZ axions, which show the photons in the superconductor become heavier than typical photons in the KSVZ model but lighter than typical photons in the DFSZ model.
That is, this effect provides a method of cross-checking the axion mass and selecting the axion model in principle.
However, this low carrier density, $n_s< 10^{14} \ \mathrm{m^{-3}}$, is not realistic, and we would like to emphasize that the difficulty lies in detecting this effect.

\section{Conclusion}
	Considering the phase synchronization and the London equations in the superconductor, it is found that the time derivative of the phase $\dot{\theta}$ enhances in the superconductor, and this value is related to the axion mass $m_a$.
The London penetration depth of a Type-I superconductor was calculated from the axion electrodynamics in the presence of a time-dependent axion field.
There is a slight possibility of detecting this effect because the effect becomes more significant in superconductor with a low carrier density $n_s$.
The London penetration depth becomes shorter in the KSVZ model but longer in the DFSZ model than the typical London penetration depth.
%%ptep

\section*{Acknowledgments}
We thank M. Kuniyasu, Professor T. Asahi and K. I. Nagao for valuable discussions, and also thank T. Takahashi who teach us the useful view for our study.

%%

%%%%%%%%%%%%%%%%%%%%%%%%%%%%%%%%%%%%%%%%%%%%%%%%%%
% \bibliographystyle{ptephy}
% \bibliography{rbunken}

\end{document}